\documentclass{aa}

\usepackage{graphicx}
\usepackage{txfonts}

\usepackage{natbib}

\usepackage{soul}
\usepackage{color}
\usepackage[colorlinks=true,citecolor=blue]{hyperref}
\begin{document}

   \title{Detection of \ion{Fe}{I} and \ion{Fe}{II} in the atmosphere of MASCARA-2b using a cross-correlation method}

   \author{M. Stangret\inst{1}\fnmsep\inst{2}, N. Casasayas-Barris\inst{1}\fnmsep\inst{2}, E. Pallé\inst{1}\fnmsep\inst{2},
   F. Yan\inst{3}, A. Sánchez-López\inst{4}, M. López-Puertas\inst{4}}
   
   \authorrunning{M. Stangret et al.}

   \institute{Instituto de Astrofísica de Canarias, Vía Láctea s/n, 38205 La Laguna, Tenerife, Spain \\
             \email{mstangret@iac.es}          
             \and 
             Departamento de Astrofísica, Universidad de La Laguna, 38200 San Cristobal de La Laguna, Spain
             \and 
             Institut f\"ur Astrophysik, Georg-August-Universit\"at, Friedrich-Hund-Platz 1, 37077 G\"ottingen, Germany
             \and
             Instituto de Astrofísica de Andalucía (IAA-CSIC), Glorieta de la Astronomía s/n, 18008 Granada, Spain
             }

   \date{Received 21 January 2020; accepted 27 February 2020}

 
  \abstract
 {Ultra-hot Jupiters are gas giants planets whose dayside temperature, due to the strong irradiation received from the host star, is greater than 2200 K. These kind of objects are perfect laboratories to study chemistry of exoplanetary upper atmospheres via transmission spectroscopy.  
 Exo-atmospheric absorption features are buried in the noise of the in-transit residual spectra. 
 However we can retrieve the information of hundreds of atmospheric absorption lines by performing a cross-correlation with an atmospheric transmission model, which allows us to greatly increase the exo-atmospheric signal. 
 At the high-spectral resolution of our data, the Rossiter-McLaughlin effect and centre-to-limb variation have a strong contribution. 
 Here, we present the first detection of \ion{Fe}{I} and the confirmation of absorption features of \ion{Fe}{II} in the atmosphere of the ultra-hot Jupiter MASCARA-2b/KELT-20b, by using three transit observations with HARPS-N. After combining all transit observations we find a high cross-correlation signal of \ion{Fe}{I} and \ion{Fe}{II} with signal-to-noise ratios of $10.5 \pm 0.4$ and $8.6 \pm 0.5$, respectively. The peak absorption for both species appear to be blue-shifted with velocities of $-6.3 \pm 0.8$~km\,s$^{-1}$ for \ion{Fe}{I} and $-2.8 \pm 0.8$~km\,s$^{-1}$ for \ion{Fe}{II}, suggesting the presence of winds from the day- to night-side of the planet's atmosphere. These results confirm previous studies of this planet and add a new atomic species (\ion{Fe}{I}) to the long list of detected species in the atmosphere of MASCARA-2b, making it, together with KELT-9b, the most feature-rich ultra-hot Jupiter to date. }

   \keywords{planetary systems -- planets and satellites: individual: MASCARA2-b -- planets and satellites: atmospheres -- techniques: spectroscopic}

   \maketitle
 
%

\section{Introduction}

High-resolution spectroscopy allows the detection and characterization of planetary atmospheres, thanks to the different Doppler velocities of the Earth, the host star and the planet. In particular, hot Jupiters, with their extended atmospheres, high temperatures, and short orbital periods, are perfect laboratories to study the nature, structure and the composition of the gas giants using these techniques \citep{2018arXiv180604617B, Snellen2010}.

More recently, a new subclass of objects, the so-called ultra-hot Jupiters (hereafter UHJ), emerged from the hot Jupiters population. UHJs are located close to their parent star, and are continually exposed to intense irradiation. This increases their temperatures to extreme values causing significant differences in chemical composition between day and night-side of the atmosphere \citep{Arcangeli2018,Bell2018, helling_2019_hp7}. The established criteria to distinguish between HJ and UHJ is the temperature in the day-side of the planet, higher than 2200 K \citep{Parmentier2018} for UHJ. Many studies also point out that in the day-side of UHJ H$_2$O is not expected, contrary to the case of hot Jupiters, because of the thermal dissociation of molecules as predicted by \citet{Parmentier2018}.

Studies of UHJs atmospheres have been carried out for several planets: HAT-P-7b \citep{Armstrong2016, helling_2019_hp7}, WASP-18b \citep{Sheppard2017WASP18, Arcangeli2018, helling_wasp_18}, WASP-33b \citep{Haynes2015, Nugroho_2017_wasp33} and WASP-103b \citep{Keidberg2018}; and a variety of atomic and molecular species have been detected in their atmospheres. 

Chemical composition studies of the hottest planet known to date, KELT-9b \citep{gaudi_kelt_9b}, reveal an extended hydrogen atmosphere (Balmer H$\alpha$ line, \citet{YanKELT9}). Moreover, \citet{Hoeijmakers_2018_kelt9, Hoeijmakers_2019_kelt9} reported the detection of \ion{Cr}{II}, \ion{Fe}{I}, \ion{Fe}{II}, \ion{Mg}{II}, \ion{Na}{I}, \ion{Sc}{II}, \ion{Ti}{II} and \ion{Y}{II}, as well as evidence of \ion{Ca}{I}, \ion{Cr}{I}, \ion{Co}{I} and \ion{Sr}{II} in its atmosphere. Additionally \citet{Cauley_2019_kelt9} detected in this planet \ion{H}{$\beta$} and \ion{Mg}{I} triplet. These atmospheric species have also been detected in the cooler UHJs WASP-18b \citep{Arcangeli2018}, WASP-33b \citep{von_essen_2018-wasp-33}, WASP-12b \citep{Jensen2012Ha}, WASP-76b \citep{seidel-2019-wasp-76} and MASCARA-2b \citep{Casasayas2018, Casasayas2019}.

In this paper, we present the study of MASCARA-2b \citep{MASCARA22017Talens} also known as KELT-20b \citep{2017Lund}. The planet orbits a fast rotating  A-type star ($v \sin{i}=116$~km\,s$^{-1}$), in 3.47 days, at a distance of 0.0542 AU. Because of the strong stellar irradiation, the equilibrium temperature of the planet is 2260~K. All stellar and planetary parameters are provided in Table~\ref{Tab:parameters}. This planet's atmosphere was recently studied using high-dispersion transit spectroscopy by \citet{Casasayas2018, Casasayas2019}, where they detected \ion{Ca}{II}, \ion{Fe}{II}, \ion{Na}{I}, \ion{H}{$\alpha$}, \ion{H}{$\beta$} and indications of \ion{H}{$\gamma$} and \ion{Mg}{I}. Here, we analyse the same three transit observations of HARPS-N used in \citet{Casasayas2019}, using the cross-correlation method. The detection of the same species via a different technique gives confidence on previous results, but also allows us to probe for other atomic/ionic species that might not have strong enough signal in individual lines to be detected in transmission. We are focusing on \ion{Fe}{I} and \ion{Fe}{II} to study the differences between the neutral and ionised species of the same atom. \ion{Fe}{I} and \ion{Fe}{II} both consist of long list of absorption lines which make them perfect species to study using cross-correlation.


\begin{figure*}[h]
  \includegraphics[width=\textwidth]{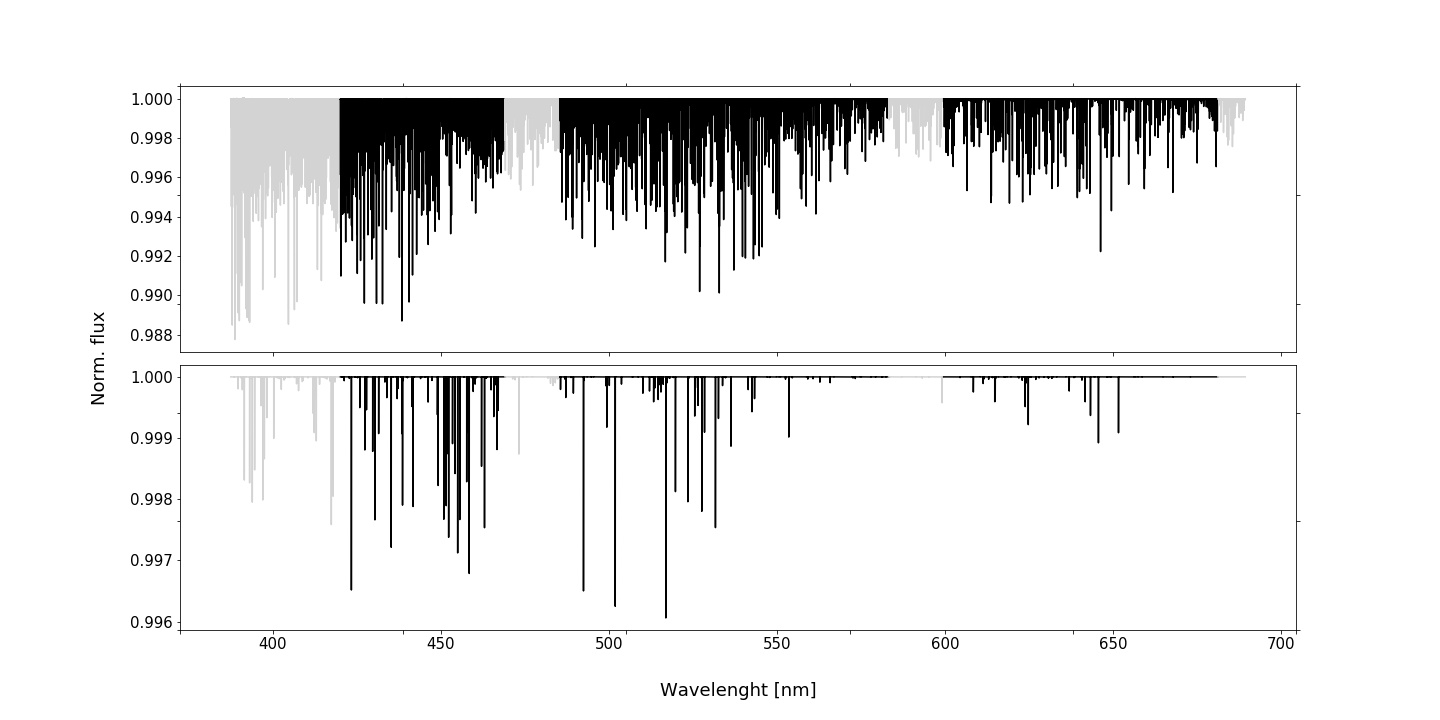}
  \caption{Synthetic models of \ion{Fe}{I} (top panel) and \ion{Fe}{II} (bottom panel). The black regions show the orders which were used in the analysis and the light gray regions show the discarded orders.}
  \label{fig:models}
\end{figure*}

\section{Observations}
 We observed three full transits of MASCARA-2b during the nights of 16 August 2017, 12 June 2018 and 19 June 2018 (Night 1, 2, and 3, from here after), using the HARPS-North spectrograph \citep{Cosentino_2012_harpsn} mounted at the 3.58-m Telescopio Nazionale Galileo (TNG) at ORM observatory (Observatorio del Roque de los Muchachos) in La Palma, Spain. During nights 2 and 3, we used the GIARPS mode which allowed us to make simultaneous observations at optical (HARPS-N) and near-infrared (GIANO-B) wavelengths. 
 
 The observation setup was as follows: for night 1 we took 90 exposures of 200s, resulting in 32 out-of-transit and 58 in transit spectra ($\phi = -0.0318$ to $+0.0350$, where $\phi$ is the planet orbital phase), with an average signal-to-noise ratio (SNR) of $51.0$; for night 2 we took 116 exposures of 200 s with 61 out of transit and 55 in transit spectra ($\phi = -0.0541$ to $+0.0393$), with an average $SNR=78.6$ and for night 3 we took 78 exposures of  300 s with 37 out of transit and 41 in transit spectra ($\phi = -0.0394$ to $+0.0441$), with an average $SNR=80.5$. During night 2, possibly because of passing clouds, the SNR for 8 spectra drastically dropped. We decided to discard them from the analysis (22:58 UT - 23:24 UT), as well as one spectra with a SNR lower than 35 taken at 01:28 UT. Thus, for night 2 we discarded 9 spectra, while for nights 1 and 3 we used all available spectra. A log of the observations is presented in Table~\ref{table:log}.

The observations were reduced using the HARPS-North Data Reduction Software (DRS, \citet{DRS1, DRS2}), version 3.7, which order-by-order extracts the spectra and performs operations such as flat-fielded using the daily calibration set.  
In the end, all the orders for each spectrum are combined into a one-dimensional spectrum. As a result we obtain one-dimensional spectra in optical wavelength given in air (380~nm - 690~nm in a step of 0.01~nm) and referred to the Barycentric rest frame.

\begin{table}
\small
\caption{Physical and orbital parameters of MASCARA-2. }             
\label{Tab:parameters}      
\centering                          
\begin{tabular}{l l l}        
\hline\hline                 
Description & Symbol & Value \\    
\hline                        
   Identifiers & - &  KELT-20, HD 185603  \\[0.1em]       
   V-band magnitude & $m_V$ & 7.6     \\[0.1em]
   Effective temperature & $T_{eff}$ &   $8980 ^{+90}_{-130}$ K  \\[0.1em] 
   Projected rotation speed    & $v \sin{i_\star}  $ &  $ 114 \pm 3$ km\,s$^{-1}$ \\[0.1em]
  Surface gravity  & $ \log{g}$ &  $4.31 \pm  0.02 $  cgs \\[0.1em]
  Metallicity  & [Fe/H] & $ -0.02 \pm 0.07$    \\[0.1em]
  Stellar mass  & $M_{\star} $ &   $1.89 ^{+0.06}_{-0.05} $  M$_\odot$ \\[0.1em]
  Stellar radius  & $ R_{\star}$ &  $ 1.60 \pm 0.06$ R$_\odot$  \\[0.1em]
  \hline
  Planet mass\tablefootmark{*}  & $M_{p} $ &  $ <3.382$ M$_J$   \\[0.1em]
  Planet radius & $R_{p} $ &  $ 1.83 \pm 0.07$ R$_J$    \\[0.1em]
  Equilibrium temperature  & $T_{eq} $ &  $2260 \pm 50 $ K  \\[0.1em]
  Surface gravity\tablefootmark{*}  & $ \log{g_p}$ &  $<3.467 $  cgs \\[0.1em]
  \hline
  Right ascension  & $...$ & $ 19^h 38^m 38.73 ^s $    \\[0.1em]
  Declination  & $...$ & $ +31^{o} 13^{'} 09.2^{''}$    \\[0.1em]
  Epoch  & $T_{c} $ & $2457909.5906 ^{+0.0003}_{-0.0002} $  BJD  \\[0.1em]
  Period\tablefootmark{*}  & $ P$ & $ 3.4741085 \pm  0.0000019$ days   \\[0.1em]
  Transit duration\tablefootmark{*}  &  $T_{14} $& $0.14898 ^{+0.00091}_{-0.00088} $  days  \\[0.1em]
  Ingress/Egress duration\tablefootmark{*}  & $ \tau$ &  $0.01996 ^{+0.00080}_{-0.00077} $ days  \\[0.1em]
  Semi-major axis\tablefootmark{*}  & $ a$ & $0.0542 ^{+0.0014}_{-0.0021} $ AU   \\[0.1em]
  Inclination\tablefootmark{*}  & $i $ & $86.12 ^{+0.28}_{-0.27} $  deg  \\[0.1em]
  Eccentricity  & $ e$ &  $ 0 $  (fixed) \\[0.1em]
  Systemic velocity  &$ v_{sys} $ & $ -21.07 \pm  0.03$ km\,s$^{-1}$    \\[0.1em]
  Projected obliquity  & $ \lambda$ & $ 0.6 \pm  4$  deg  \\[0.1em]
  Planetary RV semi-amplitude & $K_p$ & $169 \pm 10$~${\rm  kms^{-1}}$    \\[0.1em]

\hline                                   
\end{tabular}
\tablefoot{Values in the table marked with (*) were taken from \citet{2017Lund}. The rest of the parameters were taken from \citet{MASCARA22017Talens}}
\end{table}

\begin{table*}
\caption{\textbf{Summary of the transit observations of MASCARA-2b.}}             
\label{table:log}      
\centering          
\begin{tabular}{c c c c c c c c c c }     
\hline\hline       
Night & Telescope & Instrument & Date of observation & Start UT& End UT& Texp (s)& Nobs & Airmass \\ 
\hline     
    & &  &  &  &  &  & & \\ [-0.7em] 

   1 & TNG & HARPS-N & 2017-08-16 & 21:21 & 03:56 & 200 & 90& 1.089 - 1.001 - 2.089\\  
   2 & TNG & GIARPS  & 2018-07-19 & 21:27& 05:15 & 200 &116& 1.604 - 1.001 - 1.527\\
   3 & TNG & GIARPS & 2018-07-19 & 21:25& 04:23 & 300 &78& 1.012 - 1.006 - 1.903\\[0.2em]

\hline                  
\end{tabular}
\end{table*}

\begin{figure*}[h]
  \includegraphics[width=\textwidth]{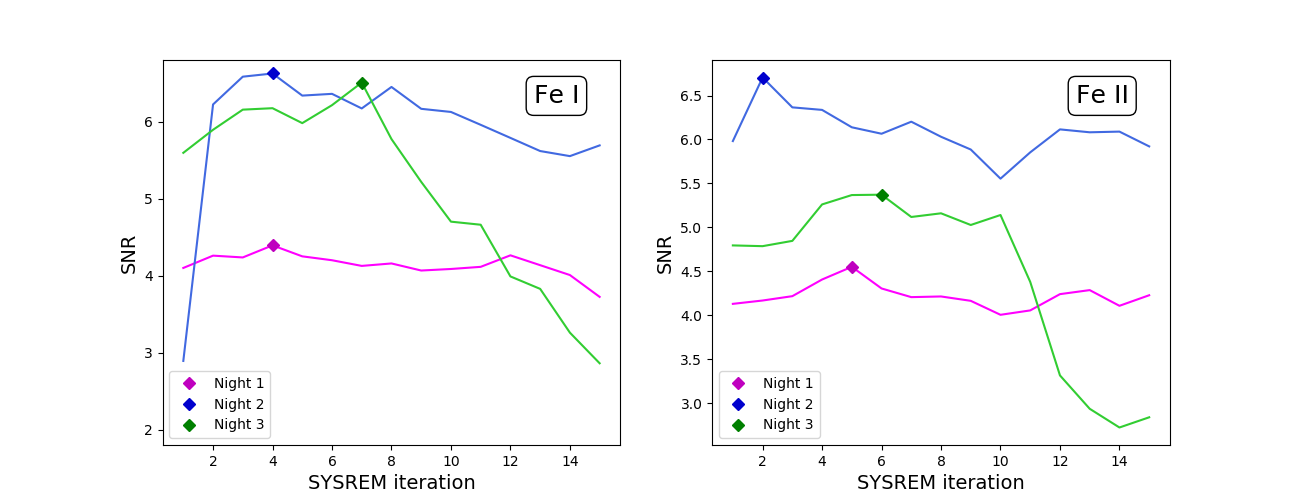}
  \caption{Maximum value of the SNR, after summing all good orders, for each SYSREM iteration and for each night; Fe I (left) and Fe II (right). The SYSREM iteration number chosen is marked with a diamond point.  }
  \label{fig:SNR_iter}
\end{figure*}

\begin{figure*}[h]
  \includegraphics[width=\textwidth]{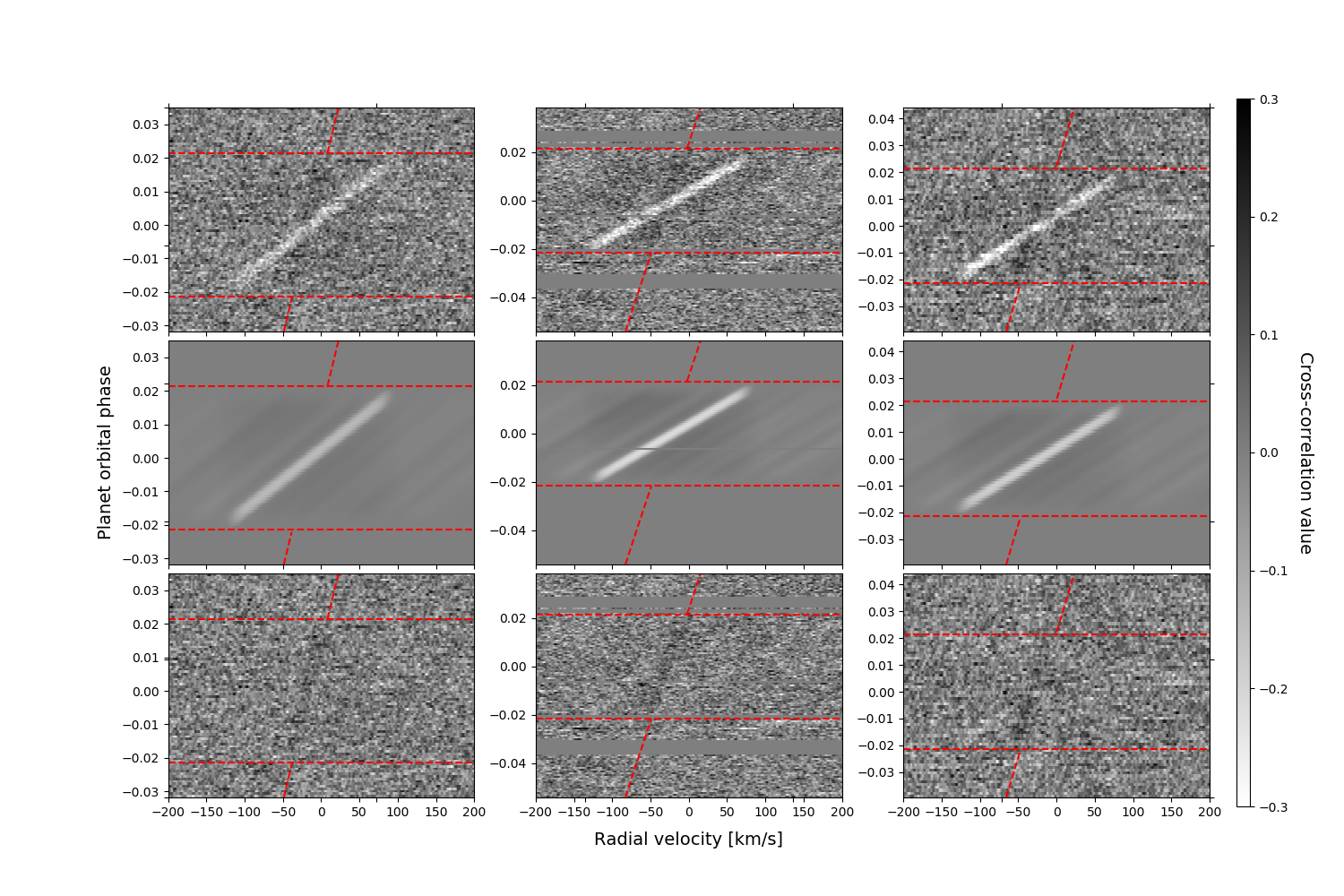}
  \caption{Cross-correlation residuals map of \ion{Fe}{I} for each night. The top three panels show the cross-correlation residuals for each night, the middle three panels shows the models of RME + CLV, and the three bottom panels shows the cross-correlation residuals after removing the effects. The horizontal red dashed line shows first and last contacts of the transit, the tilted red dashed lines show the expected planet velocities.}
  \label{iron}
\end{figure*}  

\begin{figure*}[h]
  \includegraphics[width=\textwidth]{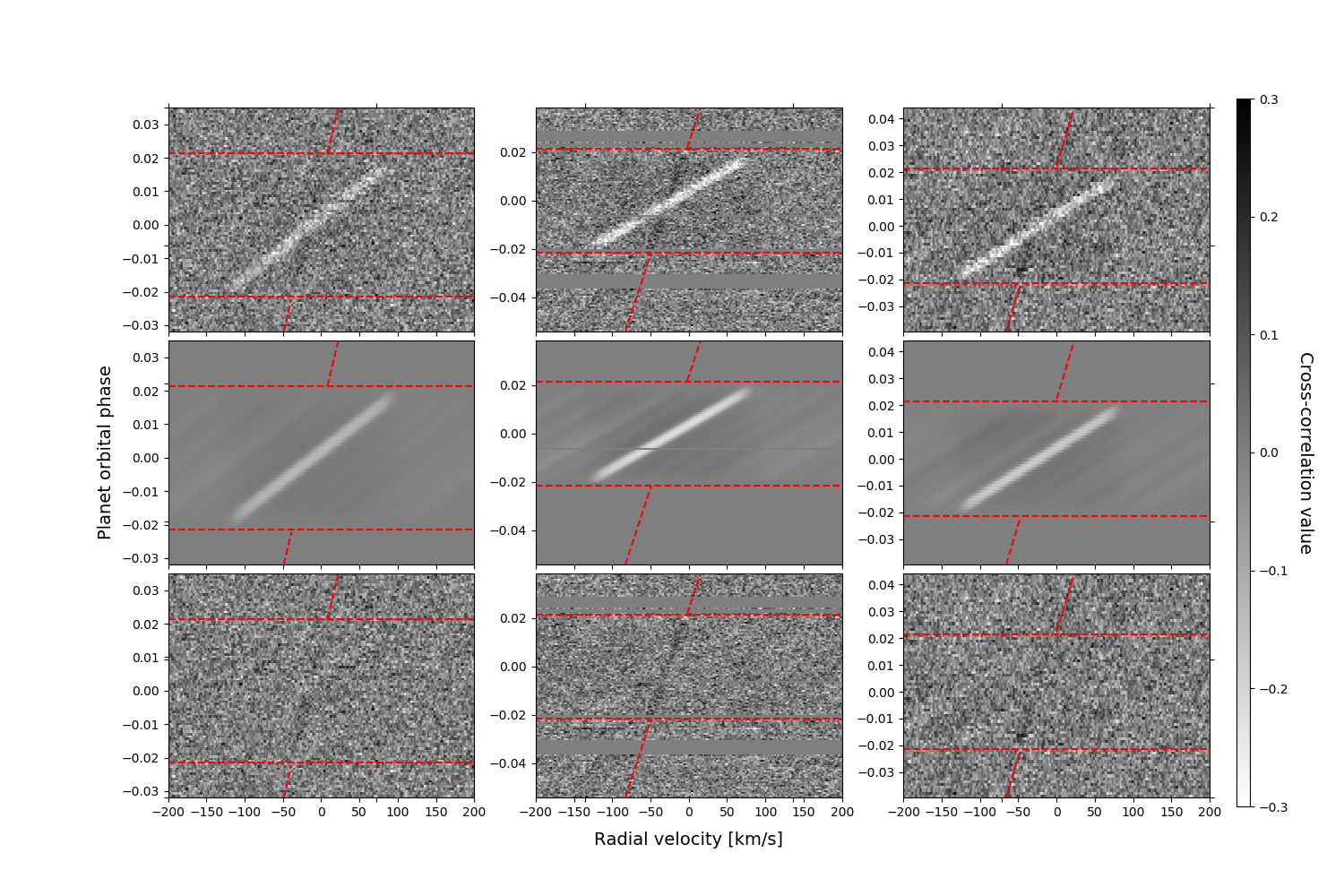}
  \caption{ The same as Fig. \ref{iron}, but for \ion{Fe}{II}}.
  \label{iron2}
\end{figure*}

\section{Methods}

\subsection{Model spectra}
Transmission spectra models of \ion{Fe}{I} and \ion{Fe}{II} used in our analysis were calculated using petitRADTRANS \citep{molliere_2019_petiRADTRANS}, which allows us to create high-resolution models of atoms and molecules at the typical temperatures of exoplanet atmospheres. To calculate the model for MASCARA-2b, we assumed a surface gravity ($\log g$) of 3.0 (corresponding to a mass of about 1.2$M_J$), considering that the actual mass of the planet is unknown. According to theoretical calculations in \citet{lothringer_barman_2019}, the upper atmospheres of UHJs orbiting A-type stars are significant hotter than the equilibrium temperatures. Therefore, we used an isothermal temperature of 4000 K, which is higher than the planetary equilibrium temperature (2260 K) but close to the simulations in \citet{lothringer_barman_2019}. The continuum level was set to 1 mbar, following \citet{yan-2019-calcium-kelt9-wasp33} and \citet{Hoeijmakers_2019_kelt9}, which state that for UHJ the continuum level is typically between 1 and 10 mbar due to H-absorption. We assumed a solar abundance and calculated the spectra of \ion{Fe}{I} and \ion{Fe}{II} separately by setting the mixing ratios of \ion{Fe}{I} and \ion{Fe}{II} to the solar Fe abundance. Finally, the spectra were convolved with the HARPS-N instrumental profile. In the case of \ion{Fe}{II} we assumed a complete ionization of Fe. The synthetic model spectra of \ion{Fe}{I}, \ion{Fe}{II} are presented in Figure \ref{fig:models}.

\subsection{Outlier rejection and normalization}

As a first step of our analysis, due to computing limitation we divide the spectrum and models into 37 chunks of 8160 pixels each. From now on, we refer to these spectral chunks in our analysis as "orders", not to be confused with the original echelle orders in the CCD. By analysing the time evolution of each pixel we removed outliers, which we define as values deviating by more than 5$\sigma$ from a fitted quadratic polynomial in time, and which are mostly related to cosmic rays. For each order independently, we normalised the spectrum fitting a quadratic polynomial. Once we have normalised the spectrum, following \citet{alonso_floriano_2018_hd189733} and \citet{sanchez_lopez_2019_hd209458} we masked all the lines where the absorption was larger that 80~\% of the flux,  which are related only to the telluric absorption, lines of studied spices are not masked. We also masked the sky emission lines where the flux was larger than 5 \% of the flux continuum.

 To correct telluric contamination we used Molecfit \citep{Molecfit1, Molecfit2}, ESO tool which fits syntetic transmission spectra to data. Following the same method as described in  \citet{Casasayas2019}. In the next step to remove the stellar signals from our data, we used SYSREM \citep{tamuz_2005_sysrem, mazeh_2007_sysrem}. This algorithm removes linear trends affecting the spectral matrix in time and wavelength.

 Each of the spectral point is weighted by its error. By iterating SYSREM we can remove most of the stellar signal and any residual telluric signals that could remain after the Molecfit corrections. As pointed in \citet{birkby_2017_sysrem} despite the fact that the signal from the planet is Doppler-shifted and much smaller in amplitude than the telluric and stellar signals, it can be also removed after a numbers of iterations. Therefore it is necessary to find the optimal number of SYSREM iterations.

 In our analysis we run SYSREM for each order independently for 15 iterations. We then tested two ways of choosing the best SYSREM iteration: using the same iteration number for both species or using different iteration number choosing the one for which the SNR of the signal is the strongest. The final results are the same within error bars, so we choose to use the optimal iteration number for each species. We found that the optimal SYSREM iterations for each night and each species are slightly different: respectively for Fe I and Fe II we used iteration 4 and 5 for night 1, 4 and 2 for night 2, and 7 and 6 for night 3. We use the same number of iterations for all orders. SNR for each SYSREM iteration for all nights and both species after summing all good orders are presented in Fig. \ref{fig:SNR_iter}.

\subsection{Cross-correlation}

For each order and each SYSREM iteration separately we cross-correlate the residuals with the models of \ion{Fe}{I} and \ion{Fe}{II}. The cross-correlation is performed in the Earth’s rest frame, using a radial velocity range of $\pm200~{\rm  km~s^{-1}}$ in steps of 0.8~km\,s$^{-1}$, which correspond to the velocity step-size of the pixels for HARPS-N. Once we retrieved the cross-correlation map, we shifted it to planet rest frame using the planet radial velocities $v_p$:

\begin{equation}
    v_p(t, K_p)= K_p \sin{2 \pi \phi (t)+v_{sys} + v_{bar}(t)}
\end{equation}

where $K_p$ is the semi-amplitude of the exoplanet radial velocity, $\phi (t)$ is the orbital phase of a planet, $v_{sys}$ is the systemic velocity and $v_{bar}(t)$ is the barycentric velocity.
To be sure that we retrieved the signal from the planet, we assumed that the $K_p$ value is unknown. We then shifted the cross-correlation map to the planet rest frame for a range of $K_p$ values, from 0 to 300~km\,s$^{-1}$, in steps of 1~km\,s$^{-1}$. We expect the maximum signal from the planetary atmosphere at the predicted values of $K_p=170$~km\,s$^{-1}$ and radial velocity 0~km\,s$^{-1}$.

In the next step, by visual inspection, we discard the orders where the telluric and stellar signals were not totally removed after applying 15 SYSREM iterations. For all of the species and nights we discarded 9 orders out of 37.

The discarded orders are presented in Figure \ref{fig:models}. For each $K_p$ value we co-added the in-transit cross-correlation values, excluding the ingress and egress data, where the spectrum could present different geometries as pointed in \citet{YanKELT9} and \citet{Salz2018He}.

As described in \citet{birkby_2017_sysrem}, \citet{brogi_2018_kpmap}, \citet{alonso_floriano_2018_hd189733} and \citet{sanchez_lopez_2019_hd209458} we checked the significance of detected signals by calculating its SNR. To do this, we divided the co-added cross-correlation values by the standard deviation away from the expected signal (0~km\,s$^{-1}$) in the region from -~50~km\,s$^{-1}$ to -150~km\,s$^{-1}$ and from 50~km\,s$^{-1}$ to 150~km\,s$^{-1}$. The results are shown in the upper panel of Figures \ref{kp_iron_3} and \ref{kp_ironII_3} for individual nights and for the combination of the three night in Figure \ref{kp_all}.

\subsection{Rossiter-McLaughlin Effect and Center-To-Limb Variation}\label{RM_CLV}

MASCARA-2 is a fast-rotating star. In consequence, we detected a strong Rossiter-McLaughlin (RM) effect, which affects our cross-correlation results (see Figure \ref{iron}). In addition to the RM effect, center-to-limb variation (CLV) also changes the stellar line profile during transit \citep{Szesla2015CLV, yan-2017-clv}. Both effects depend on the region of the stellar disk blocked by the planet. 

To retrieve the signal from the planet with high precision it is necessary to remove both of this effects from our data. For this correction, we modelled the stellar spectra over the full HARPS-N wavelength coverage, containing both CLV and RM effects on the stellar lines profiles for different planet orbital phases. To compute these stellar models we applied the methodology described in \citet{Casasayas2019}. In brief, we modelled stellar spectra for different limb-darkening angles using the line list from VALD3 \citep{VALD3} and models of Kurucz ATLAS9 computed with the Spectroscopy Made Easy tool (SME,  \citealt{SME}), as presented in \citet{yan-2017-clv}, and assuming local thermodynamical equilibrium (LTE) and solar abundance. Then, the stellar models containing the CLV and RM deformations on the stellar line profiles are calculated considering the regions of the stellar disk blocked by the planet at different orbital phases, due to the geometry of the system. For this calculation we assume 1~$R_{\rm p}$~=~1.83~$R_{\rm J}$ \citep{MASCARA22017Talens}. For the remaining system parameters, such as the spin-orbit angle ($\lambda$), $P$, $R_\star$, $i$, $a$, $e$ and $v\sin{i}$, we assume the values presented in Table~1. The residuals of these models (after dividing all modelled stellar spectra by one out-of-transit stellar spectrum) are then cross-correlated with the planetary atmospheric model following the same method applied to the observed data.

We observe that, due to different assumptions in the models calculation, such as the LTE condition, the solar abundance and the 1~$R_{\rm p}$ value (which could change due to the opacity of the atmosphere at different wavelengths as presented in \citet{SnellenChromaticRM}  and \citet{2015A&A...580A..84D}), the cross-correlated model residual maps show smaller intensity amplitudes than the cross-correlated data. To account for this more muted amplitude, we fitted the cross-correlated models to the data using Markov Chain Monte Carlo (MCMC) \citep{emcee2013PASP..125..306F}.

 Finally, because of the large uncertainties on the geometry of the system, especially in the $\lambda$ value, we shifted the RM + CLV models to the same velocity slope observed in the data, after applying the cross-correlation, within the uncertainties of $\lambda$.

\section{Results}\label{Results}

\citet{Casasayas2019} reported the detection of \ion{Fe}{II} absorption features in MASCARA-2b's transmission spectrum. Here we use the cross-correlation technique to confirm that detection, and extend it to \ion{Fe}{I}. We analyzed each night individually in search for these signatures, as well as the three nights combined.

      \begin{figure*}
   \centering
   \includegraphics[width=\textwidth]{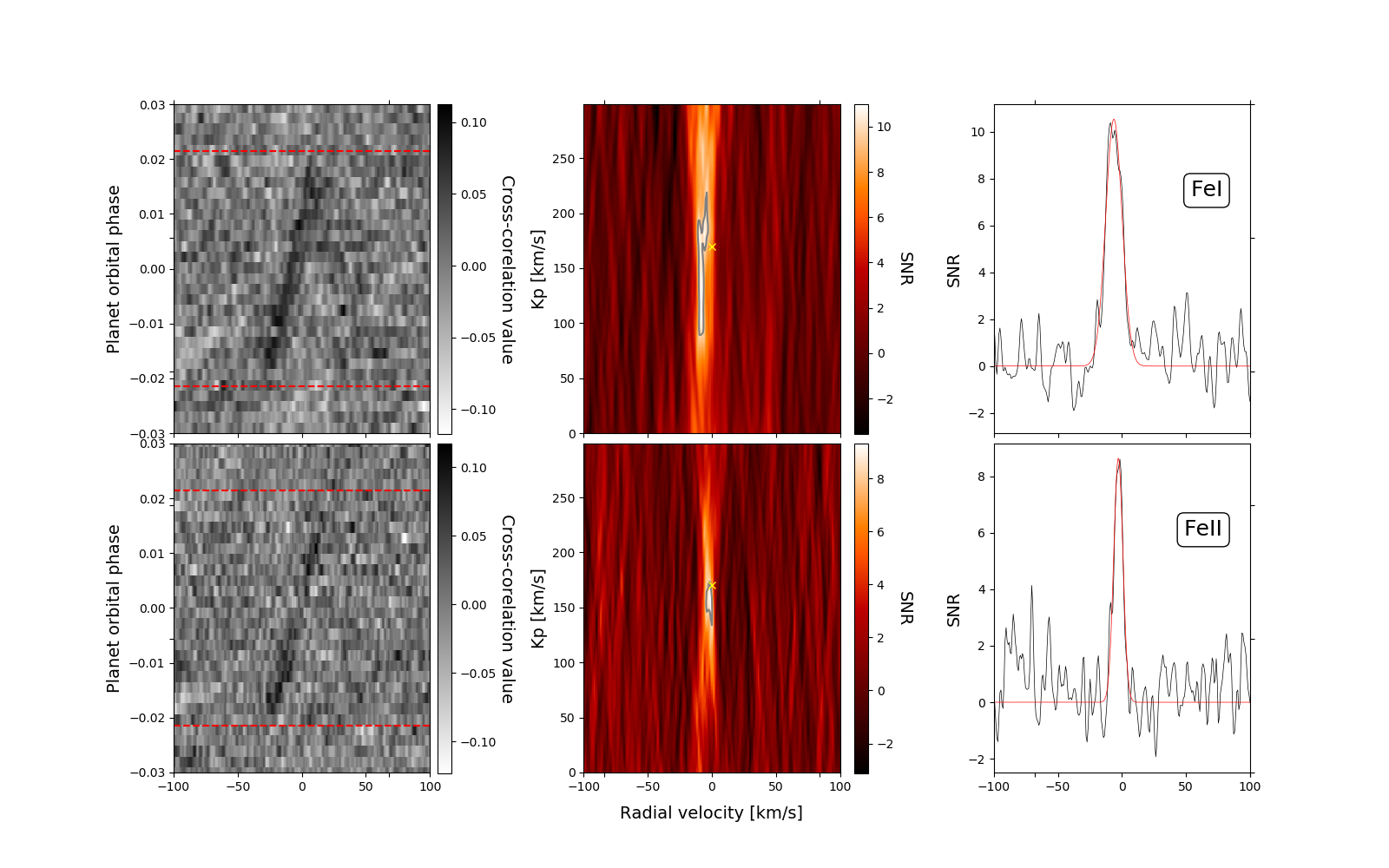}
      \caption{The cross-correlation residuals, $K_{\rm p}$ maps and SNR plots after combining the three nights for each species. The left column shows cross-correlation residual maps for \ion{Fe}{I} and \ion{Fe}{II}.  Middle column shows $K_{\rm p}$ map after cross-correlating residuals with the model of \ion{Fe}{I} and \ion{Fe}{II}. The yellow x is located at the predicted $K_{\rm p}$ value and 0~km\,s$^{-1}$. The grey contour represents the area with 1$\sigma$ from maximum signal. The right  panel shows the CCF values (black) for the predicted $K_{\rm p}$ value with the fitted Gaussian profile (red).}
         \label{kp_all}
   \end{figure*}

The cross-correlation residual maps of \ion{Fe}{I} and \ion{Fe}{II} are presented in Figures \ref{iron} and \ref{iron2}, respectively. In all figures, the three top panels show the residuals of the observations after cross-correlation. A strong signature, induced by the Rossiter-McLaughlin effect (bright tilted signal) is visible in all nights. At lower contrast, the absorption signal from the planet (dark tilted signal) is also present, with a different velocity. The three middle panels show the computed model residuals of RM + CLV, as described in section \ref{RM_CLV}. The bottom panel show the cross-correlation residual map after removing the RM + CLV model signals. The absorption features of \ion{Fe}{I} and \ion{Fe}{II} are strong enough to be detected, before and after correcting for the RM + CLV effects, for all individual nights.

As mentioned in section \ref{RM_CLV}, the cross-correlation maps (after RM + CLV correction) are shifted to the planet rest frame for different $K_{\rm p}$ values in the range 0 to 300~km\,s$^{-1}$. In Figures \ref{kp_iron_3} and \ref{kp_ironII_3} in the appendices, we present $K_p$ maps (top panels) and SNR plot for a broad range around the expected $K_p$ value for each night and each atomic transition independently. We detected the signal consistent with the exoplanet radial velocity change for each night separately at 5.1~$\sigma$, 6.2~$\sigma$ and 5.6~$\sigma$ for \ion{Fe}{I} and 4.2~$\sigma$, 6.2~$\sigma$ and 4.3~$\sigma$ for \ion{Fe}{II}.

To obtain smaller uncertainties in the $K_{\rm p}$ determination, and higher SNR, we co-add the in-transit data for each $K_{\rm p}$ value from the three nights. In Figure \ref{kp_all} we present the $K_{\rm p}$ maps and SNR plots for expected $K_{\rm p}$ value separately for \ion{Fe}{I} and \ion{Fe}{II}. In the panels, where we present the $K_{\rm p}$ maps, the yellow cross is located at the expected $K_{\rm p}$ value and 0~km\,s$^{-1}$. The grey contour represents the area with  1$\sigma$ errorbars from maximum value of SNR. We can clearly see that the signal with the highest SNR is located near the expected $K_{\rm p}$ value indicating the presence of \ion{Fe}{I} and \ion{Fe}{II} in the atmosphere of MASCARA-2b. The right panels present the SNR plots for a broad range around the expected  $K_p$ value. For \ion{Fe}{I} we detect a signal at 10.5~$\sigma$ and for \ion{Fe}{II} at 8.6~$\sigma$. In the SNR plot of \ion{Fe}{II}, we can see a slope along the negative values of radial velocities, which is related to a non-perfect removal of the RME model in this region; however it does not affect the final results. All the values of SNR, $K_{\rm p}$ and $v_{wind}$ for each night and species, as well as for the combination of the three nights, are presented in Table \ref{table:values_obtained}.

For all two models, the cross-correlation peaks appear to be blue-shifted: $-6.3 \pm 0.8$~km\,s$^{-1}$ for \ion{Fe}{I} and $-2.8 \pm 0.8$~km\,s$^{-1}$ for \ion{Fe}{II}, which suggests the presence of strong winds in the atmosphere of MASCARA-2b. This is consistent with the previous detection, and the corresponding wind velocity, derived by \citet{Casasayas2019}, a detection of \ion{Fe}{II} with $v_{wind}=-2.8 \pm 0.8$~km\,s$^{-1}$ and blue-shifted adsorptions of \ion{H}{$\alpha$}, \ion{H}{$\beta$}, \ion{H}{$\gamma$}, \ion{Ca}{II} and \ion{Na}{I}. 

It must be noted that in both works, the value for the systemic velocity of MASCARA-2 was taken from \citet{MASCARA22017Talens}, which report values of $-21.07 \pm 0.03 $~km\,s$^{-1}$ (obtained by fitting Rossiter-McLaughlin Effect) and $-21.3 \pm 0.4$~km\,s$^{-1}$ (obtained by fitting radial velocities). However, \citet{2017Lund} report a systemic velocity for MASCARA-2b of $-23.3 \pm 0.3$~km\,s$^{-1}$ (obtained by fitting radial velocities). After considering those differences and their uncertainties, the cross-correlation peaks can be detected in the range of -7.13~km\,s$^{-1}$ to -2.87~km\,s$^{-1}$ for \ion{Fe}{I} and -4.63~km\,s$^{-1}$ to -0.37~km\,s$^{-1}$ for \ion{Fe}{II}. All these results are consistent with the previous statement of blue-shifted absorption and the presence of winds in the planet's atmosphere.

There is a significant difference between the wind velocities obtained for \ion{Fe}{I} and \ion{Fe}{II}, and at the same time, the widths of the fitted gaussians to the CCF values are significantly different (FWHM = $15.1 \pm 0.6$~km\,s$^{-1}$ for \ion{Fe}{I} and FWHM = $8.5 \pm 0.6$~km\,s$^{-1}$  for \ion{Fe}{II}). These two facts suggest that the signals come from different altitudes in the planetary atmosphere, where \ion{Fe}{II} would be located in the upper part of the atmosphere and \ion{Fe}{I} deeper, where pressure is larger causing spectral broadening of its lines. This would be in line with \citet{2020arXiv200106430G} and \citet{sing-2019-wasp-121} studies of the atmosphere of the UHJ WASP-121 b, where they discuss how \ion{Fe}{I} and \ion{Fe}{II} signals are formed at physically different regions, with \ion{Fe}{I} becoming strongly ionised into \ion{Fe}{II} in the upper atmosphere.

In our case the wind velocity of neutral \ion{Fe}{} is higher than the velocity of ionised \ion{Fe}. This result contradicts predictions from the physical models of the winds in the atmosphere of giant planets \citep{2019ApJ...883....4S}, where one expects that in the higher part of the atmosphere, where the atoms are strongly irradiated and ionised, the wind velocities would be larger. At present we do not have a physical explanation for this observation.

\citet{Casasayas2019} reported detection of \ion{Fe}{II} using transmission spectroscopy method. The detected three lines of \ion{Fe}{II} ($\lambda$=501.8 nm, $\lambda$=516.9 nm and $\lambda$=531.6 nm), are three of the strongest lines and are well-separated from the other lines in the absorption spectrum of \ion{Fe}{II}. However there was no detection on individual \ion{Fe}{I}. This was probably due to a combination of the location of strong \ion{Fe}{I} lines, combined with the SNR of the orders where they lie and the location of telluric atmospheric lines. It is in fact surprising that some lines of \ion{Fe}{II} were detected alone in transmission. This gives the cross-correlation method a big advantage over transmission, because all lines have an additive contribution to the final result. In contrast, the results obtained using the cross-correlation method with atoms or molecules which have few spectral lines is challenging. 

While our results show that cross-correlation signal for \ion{Fe}{I} is stronger than for \ion{Fe}{II}, we can not determine which species is more abundant because detected signal depends on the number and strength of the spectral lines for each species (see Figure \ref{fig:models}). Further, the correlation signal will also be affected by different telluric residuals and other astrophysical/instrumental noises, making the comparison of abundances challenging. As pointed out by \citet{alonso_floriano_2018_hd189733} and \citet{2018arXiv180604617B} the CCF is practically insensitive to changes of p-T profile at present SNR levels, meaning we can also not determine the temperatures where the lines are formed.

\begin{table*}
\caption{Results obtained in the analysis for different nights and species with 1-$\sigma$ uncertainties. }             
\label{table:values_obtained}      
\centering          
\begin{tabular}{c c c c c c  }     
\hline\hline       
Night & Atom & SNR & K$_p$ [~km\,s$^{-1}$] & v$_{wind}$ [~km\,s$^{-1}$]  \\ 
\hline                    
   1 & \ion{Fe}{I} & $ 5.1\pm  0.4$ & $ 118^{+132}_{-27}  $& $-4.9 \pm 0.8 $ \\[0.1em]  
   2 & \ion{Fe}{I} & $6.2 \pm  0.4$  & $ 128^{+109}_{-48}  $ & $ -5.9\pm 0.8$\\[0.1em]
   3 & \ion{Fe}{I} & $ 5.6\pm 0.3$ & $ 142^{+108}_{-37}  $ &$-8.7 \pm 0.8 $\\[0.1em]
   All & \ion{Fe}{I} & 10.5 $\pm$ 0.4 &  121$^{+86}_{-29}  $ & --6.3 $\pm$ 0.8 \\[0.3em]
   \hline
   1 & \ion{Fe}{II} & $ 4.2\pm 0.5 $ & $ 153^{+32}_{-36} $ & $ -1.67\pm 0.8$ \\[0.1em]  
   2 & \ion{Fe}{II} & $ 6.2\pm0.5 $ &$ 150^{+31}_{-34}  $ & $ -2.9\pm0.8 $\\[0.1em]
   3 & \ion{Fe}{II} & $ 4.3\pm 0.4 $ & $ 157^{+42}_{-35}  $ & $ -3.9\pm0.8 $\\[0.1em]
   All & \ion{Fe}{II} &  8.6 $\pm$ 0.5  &  155$^{+18}_{-19}   $ &  --2.8 $\pm$ 0.8\\[0.1em]

\hline                  
\end{tabular}
\end{table*}

\section{Conclusions}

According to \citet{helling_wasp_18} in the day-side of atmospheres of ultra-hot jupiters, due to thermal dissociation, we expect the presence of both \ion{Fe}{I} and \ion{Fe}{II}. Here we present the detection of both species in the atmosphere of the ultra-hot jupiter MASCARA-2b using the cross-correlation method. The absorption features for both species are found to be blue-shifted. This suggests that we are observing strong day-to-night winds at the planet terminator, with values which are in agreement with previous studies of the atmosphere of this planet \citep{Casasayas2018, Casasayas2019}. The strong signal from \ion{Fe}{II} in the planet terminator, where temperature is significantly lower than in the day-side, also suggests the existence of winds in the atmosphere of MASCARA-2b, which would transport ionised Fe from the day-side to the night-side of the planet.

Fe I and Fe II have been previously detected in the atmosphere of KELT-9b by \citet{Hoeijmakers_2018_kelt9, Hoeijmakers_2019_kelt9} using the cross-correlation method and by \citet{Cauley_2019_kelt9} using transmission spectroscopy.
KELT-9b, however, has an extremely high day-side equilibrium temperature ($T_{eq}= 3920~{\rm K}$), while MASCARA-2b's is much lower ($T_{eq}= 2260~{\rm K}$). At these cooler temperatures, \ion{Fe}{II} has only been detected in WASP-121b (\citet{sing-2019-wasp-121}; $T_{eq}= 2700~{\rm K}$) using near-ultraviolet transmission spectrum from Hubble Space Telescope (HST). This makes MASCARA-2b one of the most interesting exoplanets known for detailed further studies of hot Jupiter atmospheres.

%

\begin{acknowledgements}
 Based on observations made with the Italian Telescopio Nazionale Galileo (TNG) operated on the island of La Palma by the Fundación Galileo Galilei of the INAF (Istituto Nazionale di Astrofisica) at the Spanish Observatorio del Roque de los Muchachos of the Instituto de Astrofisica de Canarias.
 This work is partly financed by the Spanish Ministry of Economics and Competitiveness through projects ESP2014-57495-C2-1-R, ESP2016-80435-C2-2-R and ESP2017-87143-R.
 M. S. and N.C.B acknowledge the support of the Instituto de Astrofísica de Canarias via an Astrophysicist Resident fellowship.
 F.Y. acknowledges the support of the DFG priority program SPP 1992 "Exploring the Diversity of Extrasolar Planets (RE 1664/16-1)".
 
 This work made use of PyAstronomy and of the VALD database, operated at Uppsala University, the Institute of Astronomy RAS in Moscow, and the University of Vienna.

\end{acknowledgements}

\bibliographystyle{m2.bst} 
\bibliography{m2.bib}


\onecolumn

\begin{appendix}
\section{Individual transmission spectra}

Following the same format as Figure \ref{kp_all}, we show the $K_p$ vs radial velocity maps for each individual night, together with the SNR vs radial velocity plots.

\begin{figure*}[h]
   \centering
   \includegraphics[width=\textwidth]{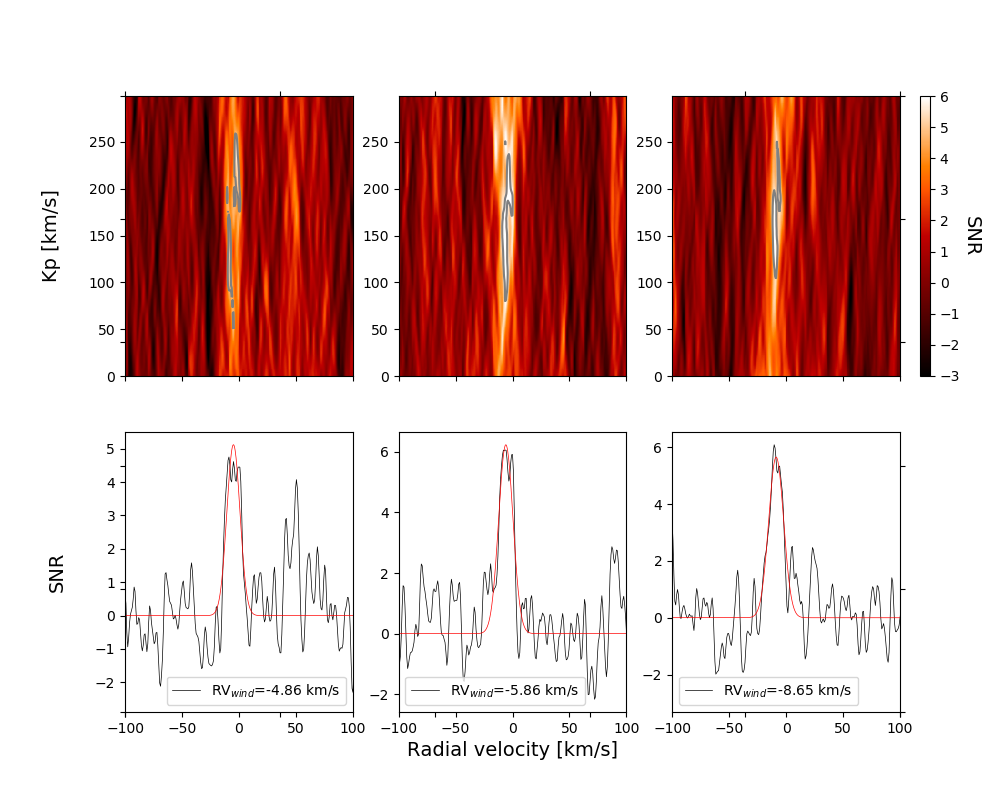}
      \caption{The $K_p$ maps and SNR plots for each of the nights separately for Fe I.  }
         \label{kp_iron_3}
\end{figure*}

  \begin{figure*}[h]
   \centering
   \includegraphics[width=\textwidth]{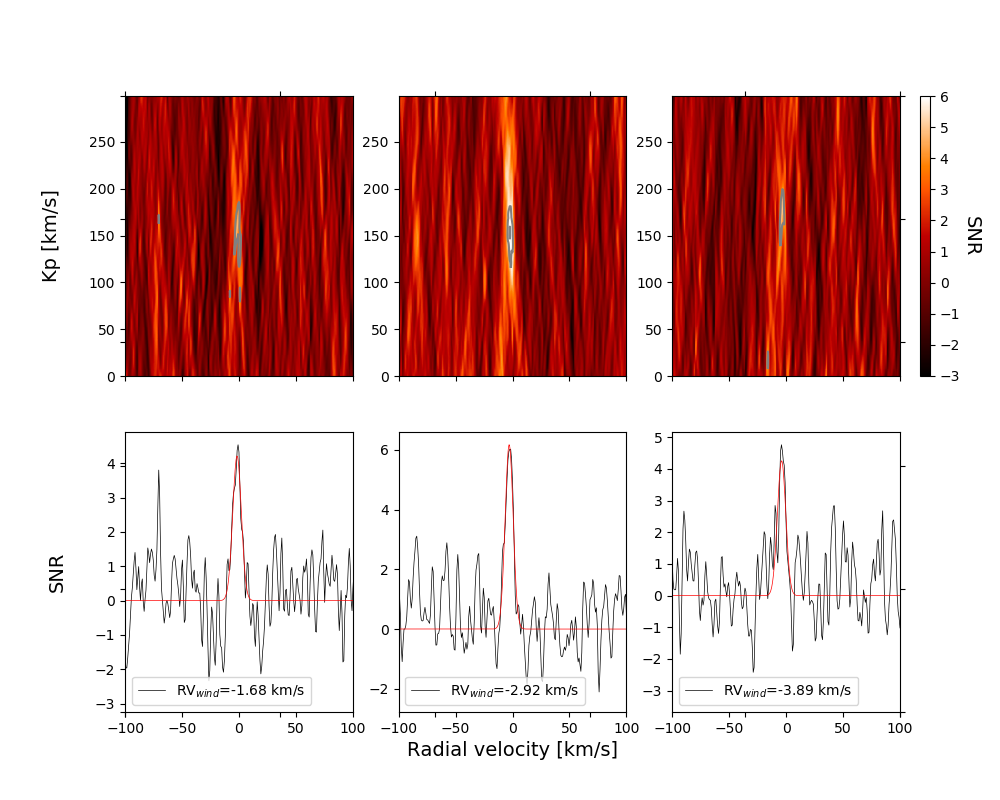}
      \caption{The $K_p$ maps and SNR plots for each of the nights separately for Fe II.}
         \label{kp_ironII_3}
   \end{figure*}

\end{appendix}

\end{document}